\pdfoutput=1
\documentclass[prd,aps,showpacs,amsmath,amssymb,
nofootinbib,unsortedaddress,twocolumn]{revtex4-1}
\usepackage{graphicx}
\usepackage{bm}

\newcommand{\be}{\begin{equation}}
\newcommand{\ee}{\end{equation}}
\newcommand{\bea}{\begin{eqnarray}}
\newcommand{\eea}{\end{eqnarray}}
\begin{document}
\title{Canny Algorithm: A New Estimator for Primordial Non-Gaussianities} 
\date{\today}
\author{Rebecca J. Danos}\email[email: ]{rjdanos@hep.physics.mcgill.ca}
\affiliation{Department of Physics, McGill University, 
Montr\'eal, QC, H3A 2T8, Canada}
\affiliation{CITA National Fellow, 
Department of Physics and Astronomy, University of Manitoba,
Winnipeg, MB, R3T 2N2, Canada}
\author{Andrew R. Frey}\email[email: ]{a.frey@uwinnipeg.ca} 
\affiliation{Dept.\ of Physics and Winnipeg Institute for Theoretical 
Physics,
University of Winnipeg, Winnipeg, MB, R3B 2E9, Canada}
\author{Yi Wang}\email[email: ]{wangyi@hep.physics.mcgill.ca} 
\affiliation{Department of Physics, McGill University, 
Montr\'eal, QC, H3A 2T8, Canada} 

\begin{abstract}
  We utilize the Canny edge detection algorithm as an estimator for
  primordial non-Gaussianities. In preliminary tests on
  simulated sky patches with a window size of 57 degrees and multipole
  moments $l$ up to 1024, we find a $3\sigma$ distinction between maps with
 local non-Gaussianity $f_{NL}=350$ (or $f_{NL}=-700$)
and Gaussian maps.  We present evidence
that high resolution CMB studies will strongly enhance the sensitivity of the
Canny algorithm to non-Gaussianity, making it a promising technique to estimate
primordial non-Gaussianity.
\end{abstract}
\maketitle

\section{Introduction}
An important emerging issue in contemporary cosmology is to search for
signatures of primordial non-Gaussianities in the Cosmic Microwave
Background (CMB). A single field inflation scenario with standard
kinetic term, slow roll potential, and standard vacuum initial
condition produces a nearly scale-invariant and nearly Gaussian
distribution of fluctuations in the CMB, while generalizations of the
simplest model, as well as alternatives to inflation, can lead to
non-Gaussianities with different sizes and shapes. Thus, a detection of
non-Gaussianity would be a significant discovery, providing strong
hints about the nature of inflationary or alternative models.

Among different types of non-Gaussianities, the most natural and well
studied one is the local shape non-Gaussianity,
\be\label{local}
\Phi(x)=\Phi_g(x)+f_{NL}\left(\Phi_g(x)^2-\langle\Phi_g(x)^2\rangle\right)
+g_{NL}\Phi_g^3(x)+\cdots\ , \ee
where $\Phi_g(x)$ is a field of Gaussian fluctuations.  In this note,
we focus on the lowest order of 
this local form non-Gaussianity, parameterized by
$f_{NL}$.

Currently there have been a number of methods to search for
non-Gaussian signatures, for example, bispectrum analysis
\cite{Komatsu:2001rj}, Minkowski functionals \cite{0067-0049-141-1-1},
and mode decomposition \cite{Fergusson:2006pr}. When applied to WMAP
\cite{Komatsu:2010fb} data with resolution of 0.21 degrees, current
constraints of non-Gaussianity are of order $-10<f_{NL}<74$
(95\% CL) from the bispectrum method. Current CMB experiments such as
Planck \cite{2010A&A...520A...1T} with resolution of 5 arcminutes, the
Atacama Cosmology Telescope \cite{Kosowsky:2004sw}
with resolution of 0.9 arcminutes, and the
South Pole Telescope \cite{Ruhl:2004kv} with
resolution of 0.25 arcminutes will result in improved constraints with
their expanded range of multipole moments and increased sensitivity
and resolution.

We will explore whether edge detection algorithms are efficient at
distinguishing non-Gaussian from Gaussian CMB skies.  In recent years,
there has emerged an interest in applying the Canny algorithm
\cite{Canny:1986:ACA}, an edge detection algorithm which searches for
steep gradients in images, to cosmological data
\cite{2008JCAP...04..015A, 2009JCAP...02..009S,2010IJMPD..19..183D,
  2010JCAP...02..033D,2010arXiv1012.3667F}.  When applied to CMB
temperature maps, the Canny algorithm selects the steep gradients in
temperature and stores them as edges.  For example, an edge map of a
temperature map in which the background possesses one temperature and
the area inside a circle possesses a different temperature would appear
as just the outline of the circle, since the edge of the circle is a
region with a steep gradient.  

Since the temperature fluctuations of Gaussian and non-Gaussian maps
have different probability distributions, locally maximal gradients
occur at different locations in the maps.  More concretely, 
\cite{Pogosyan:2009rg,Pogosyan:2011qq}
have shown that the gradients of Gaussian and non-Gaussian maps have
different probability distributions; as edges are a subset of
gradients, we expect that the edge distribution will differ between
Gaussian and non-Gaussian maps.  Heuristically, 
\cite{Pogosyan:2009rg,Pogosyan:2011qq} also
demonstrated that the number of local temperature maxima increases (or
decreases) in a non-Gaussian map as compared to a Gaussian map, along
with a corresponding decrease (or increase) in the number of local
minima.  Since one place edges will appear is along the steep
gradient between a local maximum and a local minimum, we have another
reason to expect a difference in edge statistics between Gaussian and
non-Gaussian maps.  In this paper, we take an empirical approach to
ask how sensitive the Canny algorithm is to local-form
non-Gaussianity.\footnote{It is well known that the single
  point probability distribution is not efficient enough to detect
  non-Gaussianity. However, here by considering edges, spatial
  correlation of data points are considered thus the information
  contained in edge detection is more than that in the single point
  probability distribution function.}  
To study this
question, we have used the publicly available full sky maps provided by
Elsner and Wandelt \cite{Elsner:2009md}.

In this letter, we report preliminary results testing the sensitivity
of the Canny algorithm to non-Gaussianities of the local shape in the
CMB.  We begin with a brief discussion of our simulated skies, then
review the Canny algorithm and our rough optimization of it, continue
with a discussion of our statistics and results, and conclude with
predictions related to current experiments and future simulations.

\section{Simulations}
A challenge in developing new methods for detecting CMB
non-Gaussianity is the production of simulated non-Gaussian CMB sky
maps evolved to the time of recombination with the appropriate
transfer function; this is a computationally intensive process.
Fortunately, the authors of \cite{Elsner:2009md} have provided the
spherical harmonic coefficients $a_{lm}$ for a thousand realizations
of full sky maps for both the linear and nonlinear components of the CMB
with a local form of non-Gaussianity.  That is, they provide the set
of $a_{lm}$ for both the $\Phi_g$ and $\Phi_g^2$ terms in
(\ref{local}).  These simulations include up to mulitpole moments of
$l=1024$; the spectrum of $C_l$ used in the simulations is available with
their simulations.
With these simulated maps, the work of constructing local shape
non-Gaussian maps reduces to a superposition between the Gaussian and
the non-Gaussian maps, with coefficient $f_{NL}$ in front of the
non-Gaussian maps. We anticipate future studies when simulations with
higher multipole moments and non-trivial trispectra are available.

Since our current implementation of the Canny algorithm requires the flat
sky approximation, we cut approximately 57 degree by 57 degree
Cartesian windows from these simulations. These window sizes are
compatible with the flat sky approximation. Any distortion due to the
Cartesian slicing and flat sky approximation applies equally to both the
Gaussian and non-Gaussian maps and hence does not affect our results.

As a check that our algorithm is not sensitive to differences between
independent Gaussian maps, we have compared two sets of 120 Gaussian
maps using the statistics described below.  We find that there is a 96\%
chance that two sets of maps drawn from the same distribution would 
have greater statistical differences, so they are
indistinguishable.  In addition, we have checked that the non-Gaussian
maps and Gaussian maps have the same spectrum ($C_l$).


\begin{figure}
\center
\includegraphics[width=0.42\textwidth]{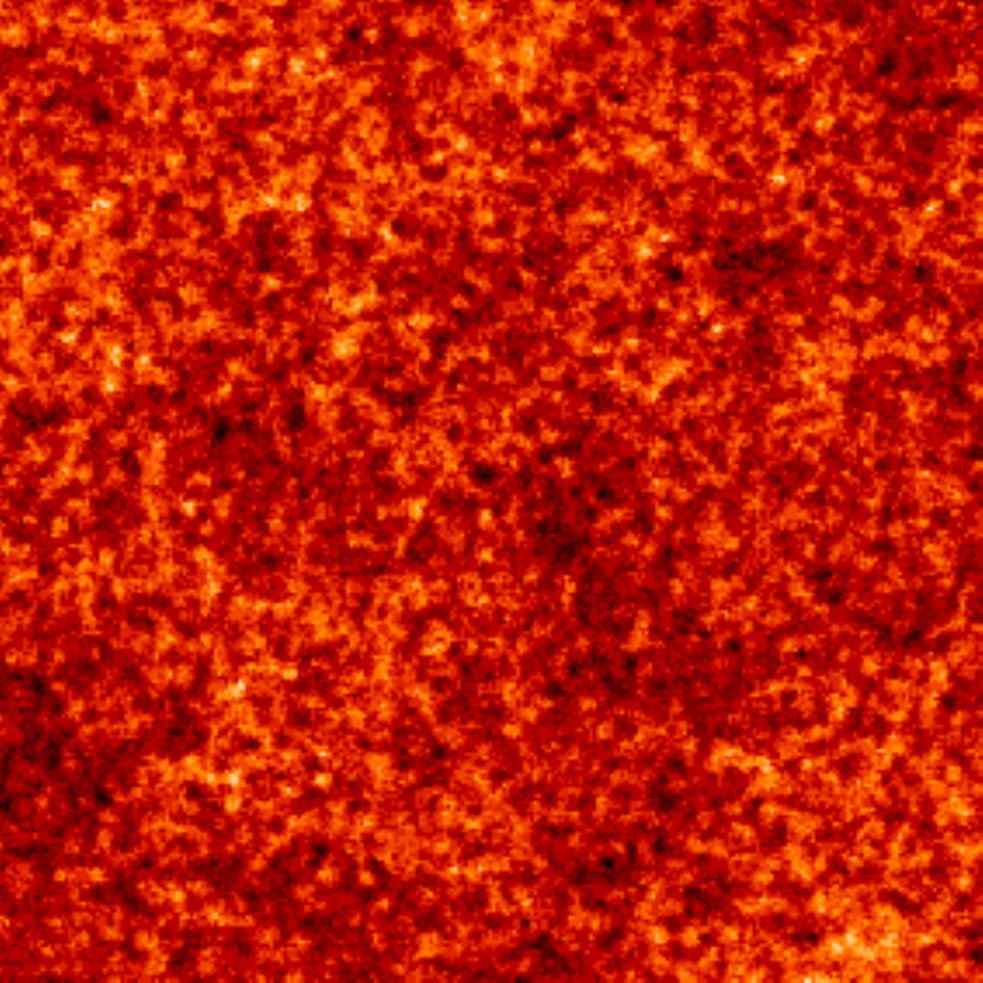}
\caption{\label{fig:gaussianmap}
A sample of the Gaussian map, in a window of 57 degree by 57 degree
sky. }
\end{figure}

\begin{figure}
\center
\includegraphics[width=0.42\textwidth]{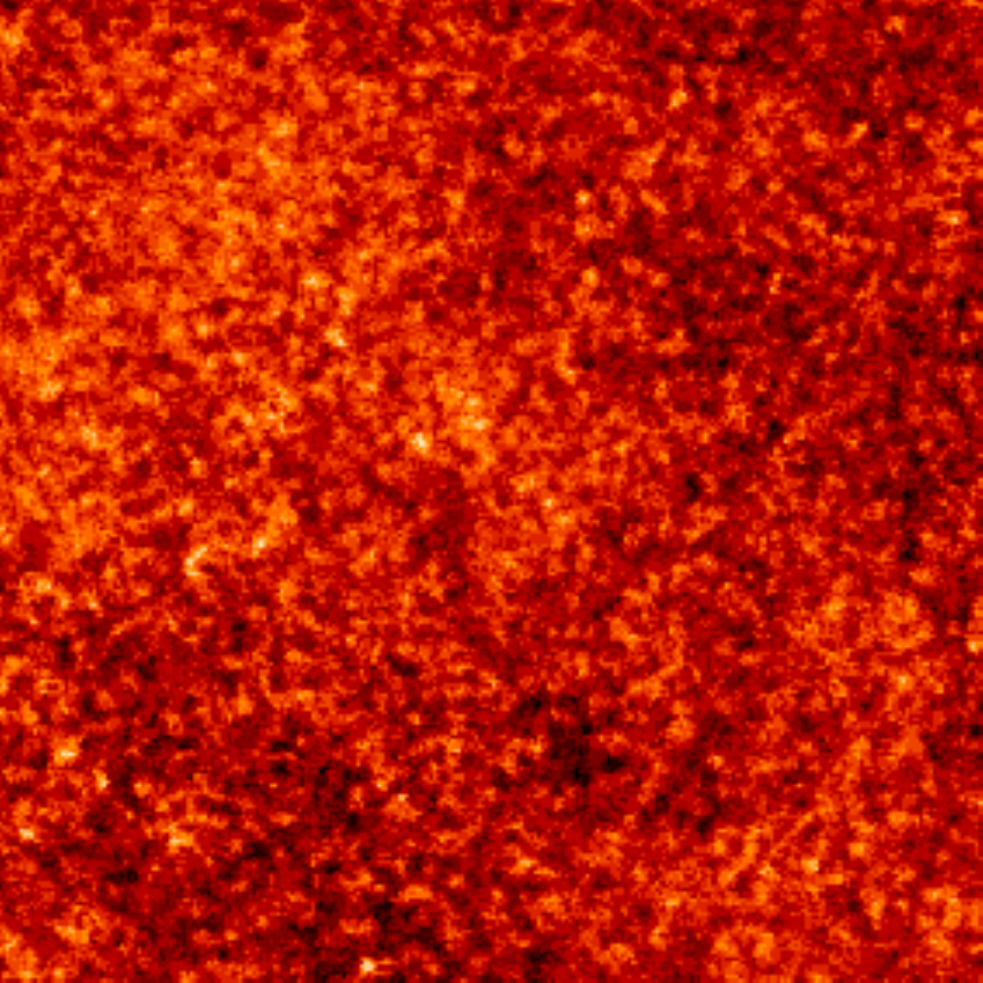}
\caption{\label{fig:ngmap} A sample of the non-Gaussian map with
  $f_{NL}=1000$, in a window of 57 degree by 57 degree sky.  }
\end{figure}

\section{Review of Canny Algorithm}
We refer the reader to \cite{2010IJMPD..19..183D} for a complete
description of our implementation of the Canny algorithm.  In brief,
we search for local maxima in the gradient
of the map along the direction of the gradient, in the following
steps: 
\begin{enumerate} 
\item Convert the temperature map to a gradient map, 
recording magnitude and direction of the derivative at each pixel.

\item Scan the map along eight directions (vertically, horizontally,
and along both diagonals) retaining 
only local maxima along the direction of the gradient.

\item Filter the remaining gradient map such that only 
gradients with magnitudes between a lower threshold $t_l$ and a
cut-off threshold $t_c$ are kept.

\item Impose an upper threshold $t_u$. The remaining points
above this threshold are considered as belonging to an edge.
Points below this threshold are considered as belonging to an edge
only if they are connected to an edge in a direction perpendicular to their
gradient.

\item Count and store the numbers and lengths of edges to
perform statistical analysis.
\end{enumerate}
The thresholds are measured in units of a maximal gradient $G_m$, which 
we define as the lesser of the averages of the maximum gradient magnitude
for all the Gaussian and non-Gaussian maps.  Again, we refer the reader to
\cite{2010IJMPD..19..183D} for more details.

One significant change made for the above standard implementation
refers to Appendix A of \cite{2010IJMPD..19..183D}, which removes
doubles of locally maximal gradients.  In choosing which doubled pixel
to discard, instead of choosing the pixel with lower temperature,
which artificially introduces more edges with lower temperature
fluctuations, we chose the pixel with the lower absolute value of the
temperature fluctuation.

In addition, we implemented a routine to optimize the thresholds used.
We sampled ten values of the threshold parameters and minimized the
quadratic fit to the probabilities for $f_{NL}=1000$ (see the
description of our statistics below).  The final thresholds used were
$t_c=3.09414$, $t_u=0.257424$, and $t_l=0.104205$.

\section{Statistics and Results}
To differentiate statistically between sets of Gaussian and non-Gaussian
simulated images, we applied the same tests as described in 
\cite{2010IJMPD..19..183D} for detection of cosmic strings.  In brief, for
each CMB map (a window of 500 pixels on a side), we divided all the edges
into bins by edge length; all edges longer than a determined maximum length
$k$ were binned with that length.  Then we found the distribution of all the
windows within each bin for both Gaussian and non-Gaussian maps.  
At that point, we apply the Student t-test to the two distributions in 
each bin.  From the t-test we obtained
the $p$-values, which give probability information,
which we then combined using the Fisher combined probability test,
\be
\chi_{2k}^2 = -2\sum_{l=1}^k \ln(p_l)\ ,
\ee
to compute the total $\chi^2$ separating the two sets of maps, which follows
the $\chi^2$ distribution with $2k$ degrees of freedom.
We then find the probability that the two sets of CMB maps
could have that value of $\chi^2$ (or larger) if they were 
drawn from the same larger distribution of maps.  In
the following, we use an optimal value of $k=2$.

\begin{figure}
\center
\includegraphics[width=0.49\textwidth]{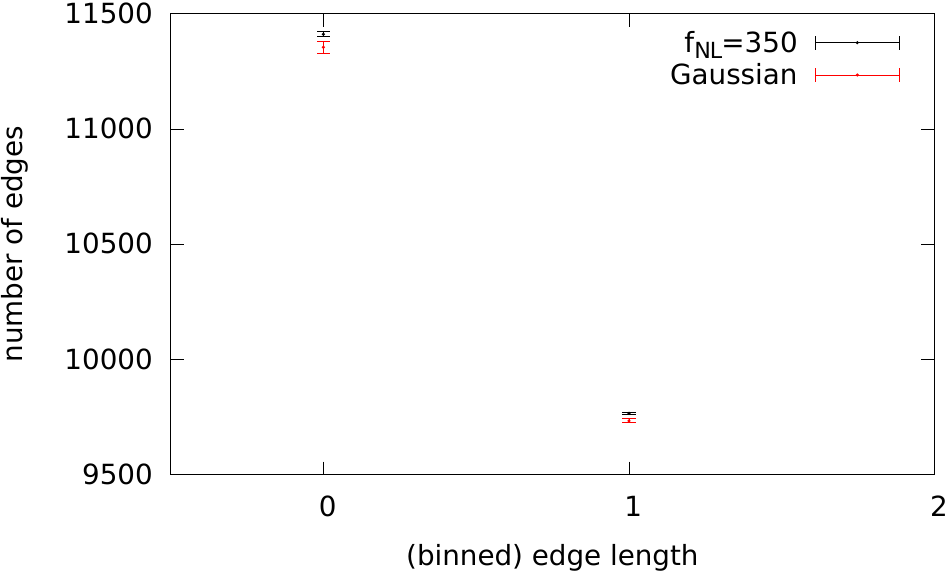}
\caption{\label{fig:fnl350}
Edge statistics for $f_{NL}=350$. The blue (upper) dot represents the 
non-Gaussian simulations, and the red (lower) dot represents the Gaussian
simulations.  Error bars are the standard error of the mean (1$\sigma$).
}
\end{figure}

\begin{figure}
\center
\includegraphics[width=0.49\textwidth]{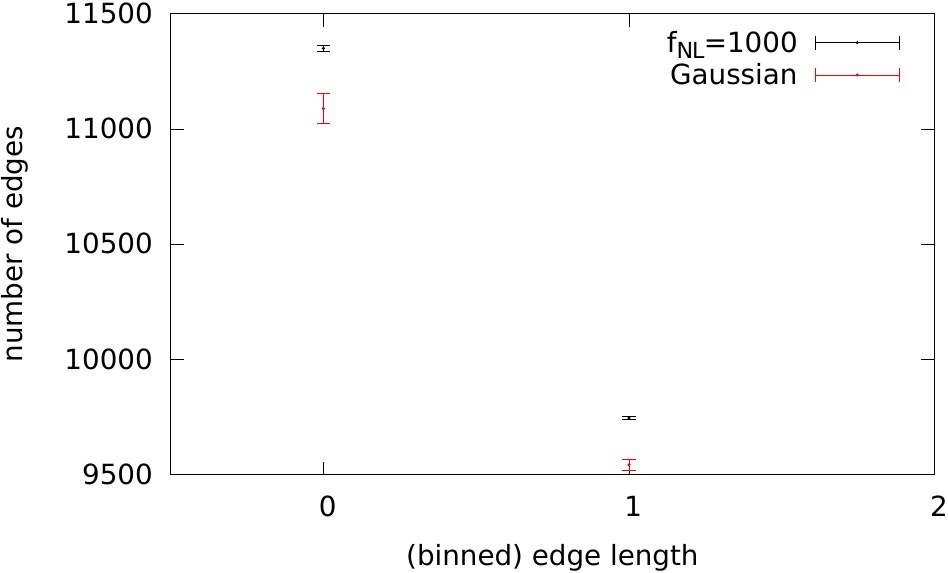}
\caption{\label{fig:fnl1000}
Edge statistics for $f_{NL}=1000$. The blue (upper) dot represents the 
non-Gaussian simulations, and the red (lower) dot represents the Gaussian
simulations.  Error bars are the standard error of the mean (1$\sigma$).
}
\end{figure}

We applied the Canny algorithm and statistical analyses to 120 windows of
approximately 57 degrees (500 pixels) per side and an $f_{NL}=350$.  Our
results indicate that there is a 0.1\% probability that the non-Gaussian
simulations and Gaussian simulations would have such a large value of 
$\chi^2$ if drawn from the same distribution, 
which would constitute a $3\sigma$ detection.  We find similar statistical
significance distinguishing Gaussian simulations and simulations with
a negative non-Gaussianity parameter of $f_{NL}=-700$.
The statistics for the comparison of $f_{NL}=350$ and Gaussian simulations
 are plotted in figure
\ref{fig:fnl350}, showing the distribution in each bin. 
As a comparison, statistics for $f_{NL}=1000$ are also shown in figure
\ref{fig:fnl1000}.  An astute reader will note that the Gaussian simulations
give different edge counts in the two figures; this is because they are 
compared to simulations with different amounts of non-Gaussianity, resulting
in different values of the parameter $G_m$ defined above.

The reader will also note that our $3\sigma$ detection level is 
asymmetric in $f_{NL}$; that is, it seems possible to distinguish smaller
$|f_{NL}|$ from Gaussian maps if the sign of $f_{NL}$ is positive.  We 
have checked that this is not due to a bias in the algorithm by reversing
the sign of all temperature fluctuations on a set of maps; these maps give
edge counts that are indistinguishable from the original set of maps.
It appears that the asymmetry is an intrinsic property of local-type 
non-Gaussianity due to the nonlinearity of the fluctuations.

As the Canny algorithm is designed to detect edges, it may be possible to
optimize it for the study of primodial non-Gaussianity (for example,
by finding new statistics to differentiate Gaussian and non-Gaussian
skies).  However, there are also prospects for improvement even with the
un-modified algorithm and statistics presented here.  For example, 
our method is 
sensitive to the number of pixels in each window.  A comparison of the same
non-Gaussian and Gaussian simulations in windows with 200 pixels per side
yields a probability of 16.3\%.  Therefore, we expect that the better
resolutions offered by future simulations and experiments such as Planck,
ACT, and SPT should allow greater sensitivity.  

In contrast, our results
were less sensitive to the number of windows considered. 
When we applied the
Canny algorithm to 30 non-Gaussian realizations with $f_{NL}=350$ and 
120 Gaussian realizations
(we have no upper constraints to the number of Gaussian realizations we can
produce), we obtained probabilities of 4\% that the maps were drawn from the
same distribution.  
In addition, we found that reducing the number of Gaussian simulations
(to 30 Gaussian and 30 non-Gaussian simulations) resulted in
probabilities of 17\% for $f_{NL}=350$.  Therefore, we conversely expect
that increasing the number of Gaussian simulations could improve our
results (concominant with improved resolution) due to the fact that the
standard error of the mean decreases with increasing sample size.  
This offers another
potential avenue to improve the Canny algorithm's sensitivity to
local-type non-Gaussianity.

\section{Conclusion}
We have shown that applying the Canny algorithm to segments of full
sky Gaussian and non-Gaussian maps can differentiate the two sets of
maps at the $3\sigma$ level down to $f_{NL}=350$ or $f_{NL}=-700$ (note
that current observational limits are quoted at the 95\% confidence level).  
Since tests of our
application greatly improved for 500 pixels per window side compared
to 200 pixels per window side, we anticipate that implementation of
this algorithm on high resolution data should dramatically improve our 
results.  For example,
SPT \cite{Ruhl:2004kv} will provide 
data with up to 2400 pixels per 10 degree side.
In particular, our tests
indicated that larger pixel numbers were more relevant to
substantially improved results than larger numbers of images.  For
this reason, application of the Canny algorithm 
to Planck \cite{2010A&A...520A...1T}, ACT \cite{Kosowsky:2004sw},
and SPT \cite{Ruhl:2004kv} data should also be very promising for
detection of primordial non-Gaussianities.

In the present note, we considered the local type bispectrum as the form
of non-Gaussianity of interest. However, our method is general and can be
performed on non-Gaussian maps with other types of non-Gaussianity,
namely bispectra with other shapes as well as a nontrivial trispectrum. 
We especially 
expect our estimator to be sensitive to the local shape
trispectrum, where the Gaussian distribution is deformed by kurtosis
instead of skewness.  In this case the slope of the probability
distribution function is changed symmetrically, so more (less) edges
should be produced when the kurtosis is positive (negative) respectively. 

On the other hand, it remains unclear whether we can distinguish
different types or shapes of non-Gaussianity or the contribution from
cosmic strings. We will leave the comparison between these signals to
future work.

In our current approach, the error bars in the figures are determined
numerically by simulations of Gaussian and non-Gaussian maps. It
remains interesting to see whether these error bars could also be
determined theoretically. In \cite{Pogosyan:2011qq}, the extrema
counts for CMB maps are calculated theoretically.  It may be possible to 
obtain analytical bounds on edge number statistics. The analytical extrema
counts may also suggest new types of statistical analysis to perform on
edge maps.

Another important issue that we have not addressed in the present note
is to add noise to the simulated maps. By adding noise, according to
the sensitivities of WMAP, Planck, SPT or ACT, we could tell what
value of $f_{NL}$ could be detected in the corresponding
experiments using the Canny algorithm. We hope to address this issue
in the future.

Finally, the Canny algorithm is originally developed to detect edges
instead of non-Gaussianity. Thus, although we have shown that the
algorithm is sensitive to non-Gaussianity, we expect there is considerable
potential to optimize the algorithm, such as through the identification of
a new statistic to distinguish Gaussian and non-Gaussian maps.  Therefore,
we are optimistic about the potential of the Canny algorithm for development
as an estimator of non-Gaussianity in the CMB. 

\vspace{.5cm}

\begin{acknowledgments}
We would like
to thank R.~Brandenberger, G.~Holder, L.~LeBlond, and A.~ van Engelen
for useful discussions.  RJD thanks the University of Winnipeg Department 
of Physics for hospitality during the completion of this work, as well
as K.~Dasgupta and M.~Venditti for their support.

\end{acknowledgments}


%

\end{document}